\documentclass[prl,twocolumn,showpacs,amsmath,amssymb,superscriptaddress]{revtex4}
\usepackage{pifont,amssymb,epsf,graphicx}
\begin{document}
\title{Nesting-induced Local Electronic Patterns around a Single As (Te, Se) Vacancy in Iron-based Superconductors}

\author{Degang Zhang}
\affiliation{College of Physics and Electronic Engineering, Sichuan Normal University,
Chengdu 610101, China}
\affiliation{Institute of Solid State Physics, Sichuan Normal
University, Chengdu 610101, China}
\author{Zhihai Liu}
\affiliation{College of Physics and Electronic Engineering, Sichuan Normal University,
Chengdu 610101, China}
\author{Jianming Ma}
\affiliation{College of Physics and Electronic Engineering, Sichuan Normal University,
Chengdu 610101, China}
\author{Jiangshan Liu}
\affiliation{College of Physics and Electronic Engineering, Sichuan Normal University,
Chengdu 610101, China}
\author{Zhengwei Xie}
\affiliation{College of Physics and Electronic Engineering, Sichuan Normal University,
Chengdu 610101, China}
\affiliation{Institute of Solid State Physics, Sichuan Normal
University, Chengdu 610101, China}
\author{C. S. Ting}
\affiliation{Texas Center for Superconductivity and Department
of Physics, University of Houston, Houston, Texas 77204, USA}

\begin{abstract}
The local electronic states around a single As (Te, Se) vacancy are investigated
in order to shed light on the role of ligands in a series of iron-based superconductors.
Such a vacancy can produce a local hopping correction ranging from $-0.22$ eV to 0.12 eV and
always induce two in-gap resonance peaks in the local density of states
(LDOS) at the fixed symmetrical bias voltages, which are rather robust and irrelevant to
the phase of superconducting order parameter. The LDOS images
near the defect predominantly possess $0^o$ and $45^o$ stripes.  These energy-dependent
charge modulations created by quasiparticle interference are originated in the
nesting effect between the inner (outer) hole Fermi surface around $\Gamma$ point and
the inner (outer) electron Fermi surface around $M$ point.

\end{abstract}

\pacs{71.10.Fd, 71.18.+y, 71.20.-b, 74.20.-z}

\maketitle

The mechanism of high temperature superconductivity has been one of the
great challenges in the condensed matter physics community since the discovery of
the cuprates in 1986 [1]. A series of high $T_c$ cuprate superconductors commonly
possesses layered crystal structures consisting of the conducting ${\rm CuO}_2$ planes
separated by the other elements and oxygen layers. The ligand O ions in the
${\rm CuO}_2$ planes just locate on the Cu-Cu bonds and are believed to play
an important role in forming superconductivity. The surface effects of the
cuprates can be neglected due to the positions of the O ions.
However, it is difficult to
evaluate the impact of the O ions on the electronic states in the ${\rm CuO}_2$ planes
due to the lattice distortion or in-plane disorders. Fortunately, new family
of high $T_c$ superconductors, i.e. iron-based superconductors, was found in 2008 [2-7].
The iron-based superconductors also have a layer crystal structure
and each unit cell contains two Fe ions (A and B)
and two As (Te, Se) ions (A and B) (see Fig. 1). The high temperature superconductivity
is originated in the electron pairing in the Fe-Fe plane by doping electrons or holes.
The ligand As (Te, Se) ions A and B are located just below and above
the center of each face of the Fe square lattice, respectively, rather than in the
conducting plane. Such a crystal structure
provides us an excellent platform for exploring the ligand effects on the
electronic states, which can easily distinguish from the disorders
in the conducting plane. Obviously, the impacts of As (Te, Se)
ions A and B in the surface layer of the iron-based superconductors on 
the local density of states (LDOS)
are inequivalent due to their different environments. It is known that
the experimental results observed by both angle resolved photoemission 
spectroscopy (ARPES) and scanning tunneling microscopy (STM) contain 
unavoidably this kind of surface effect.

\begin{figure}
\rotatebox[origin=c]{0}{\includegraphics[angle=0,
           height=2.5in]{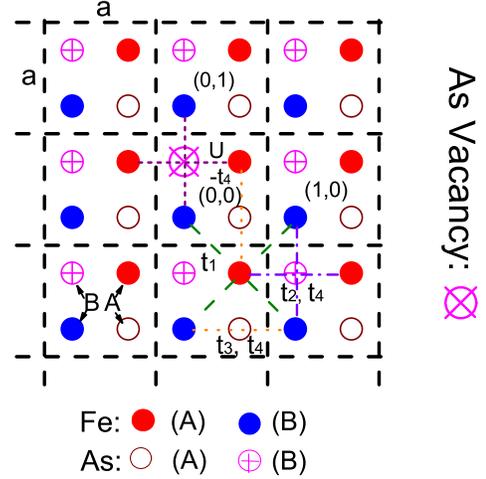}}
\caption {(Color online) Schematic of a single As (Te, Se) vacancy
in the Fe-As (Te, Se) layer with each unit cell containing two
Fe (A and B) and two As (Te, Se) (A and B) ions. The As (Te, Se) ions A and B are
located just below and above the center of each face of the Fe
square lattice, respectively. Here,
$t_1$ is the nearest neighboring
hopping between the same orbitals $d_{xz}$ or $d_{yz}$, $t_2$ and
$t_3$ are the next nearest neighboring hoppings between the same
orbitals mediated by the As (Te, Se) ions B and A, respectively, $t_4$ is the
next nearest neighboring hopping between the different orbitals,
and $U$ is the local hopping correction to $t_1$ due to the ligand
vacancy situated at the point $(0,\frac{1}{2})$ in the Fe sublattice $B$
or the point $(-\frac{1}{2},0)$ in the Fe sublattice $A$.}
\end{figure}

In order to figure out the origin of high temperature superconductivity in iron-based
superconductors, we first understand the role of the ligand As (Te, Se) ions.
Recently, Li and Yin investigated the As vacancies on the surface of
optimally electron-doped BaFe$_{2-x}$Co$_x$As$_2$  by performing STM observations and
found a pair of LDOS peaks within superconducting energy gap [8,9].
In this work, motivated by the interesting STM experiments,
we study the influence of a single As (Te, Se) vacancy on
the LDOS in the Fe-Fe plane by employing a two-orbit four-band
tight binding model [10], which takes the asymmetric effect of the ligand
As (Te, Se) ions in the surface Fe-As (Te, Se) layer into account. 
Such an empirical model can fit
excellently the energy band structure of iron-based superconductors
and its evolution with electron or hole doping measured by ARPES experiments [11-20].
This model also explained successful a series of STM experiments
in iron-based superconductors, e.g.
in-gap impurity bound states [10,21], the negative energy resonance peak
in the vortex core [22,23], the $90^o$ domain walls and anti-phase domain walls[24-27],
the zero-energy bound state induced by the interstitial excess Fe ions[28-30], etc.,
and especially repeated the phase diagram observed by
nuclear magnetic resonance and neutron scattering experiments [31-33].

\begin{figure}
\rotatebox[origin=c]{0}{\includegraphics[angle=0,
           height=1.25in]{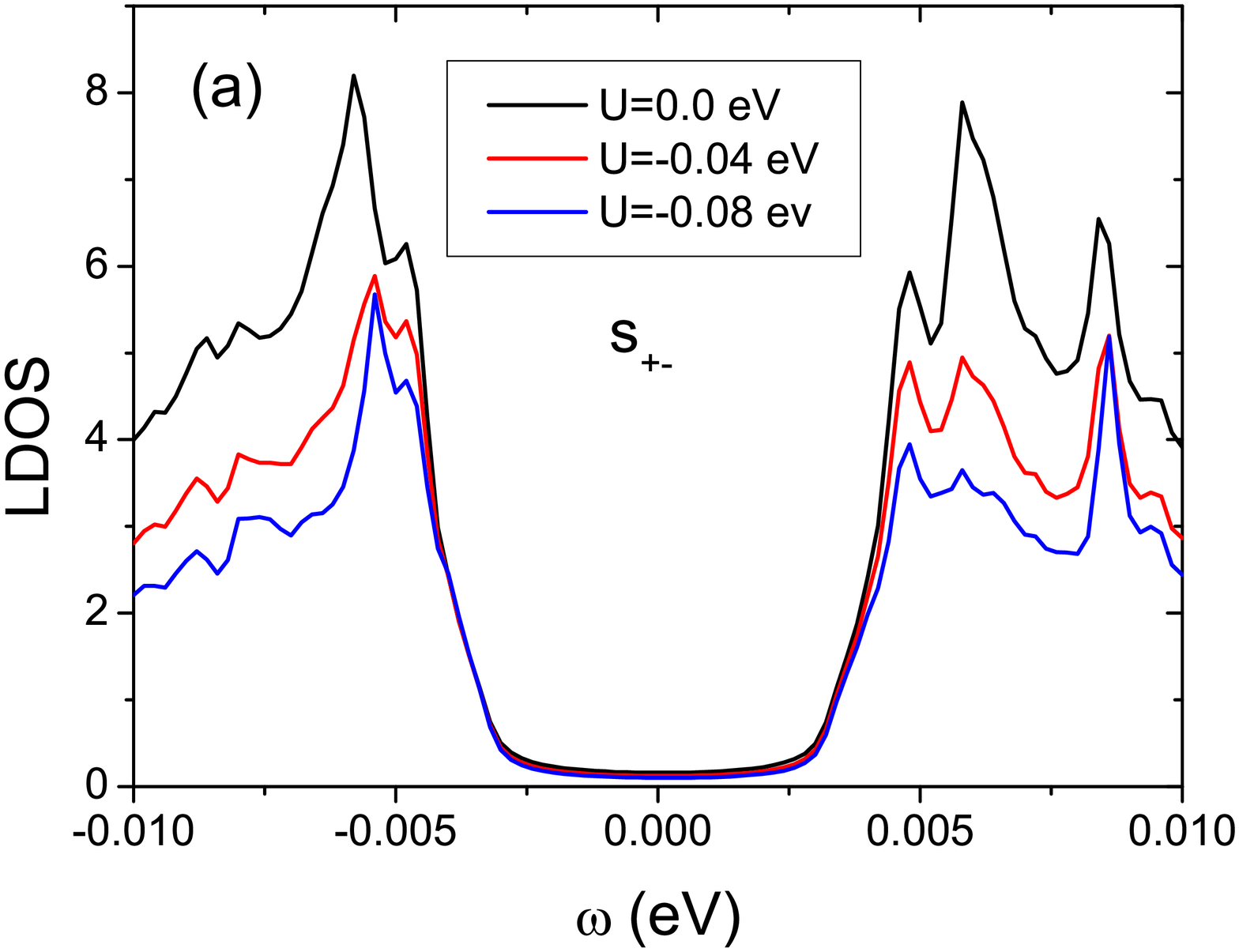}}
\rotatebox[origin=c]{0}{\includegraphics[angle=0,
           height=1.25in]{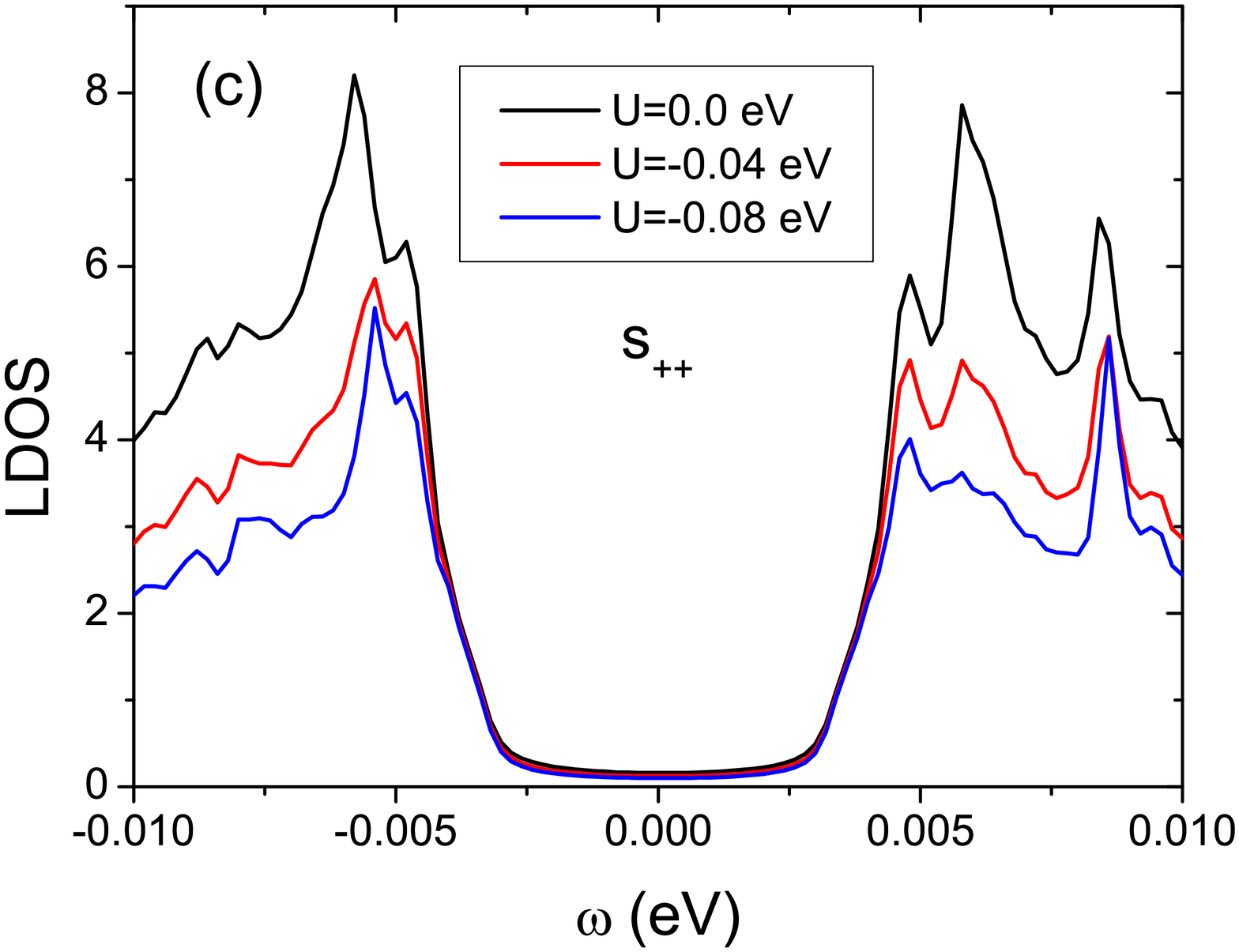}}
\rotatebox[origin=c]{0}{\includegraphics[angle=0,
           height=1.25in]{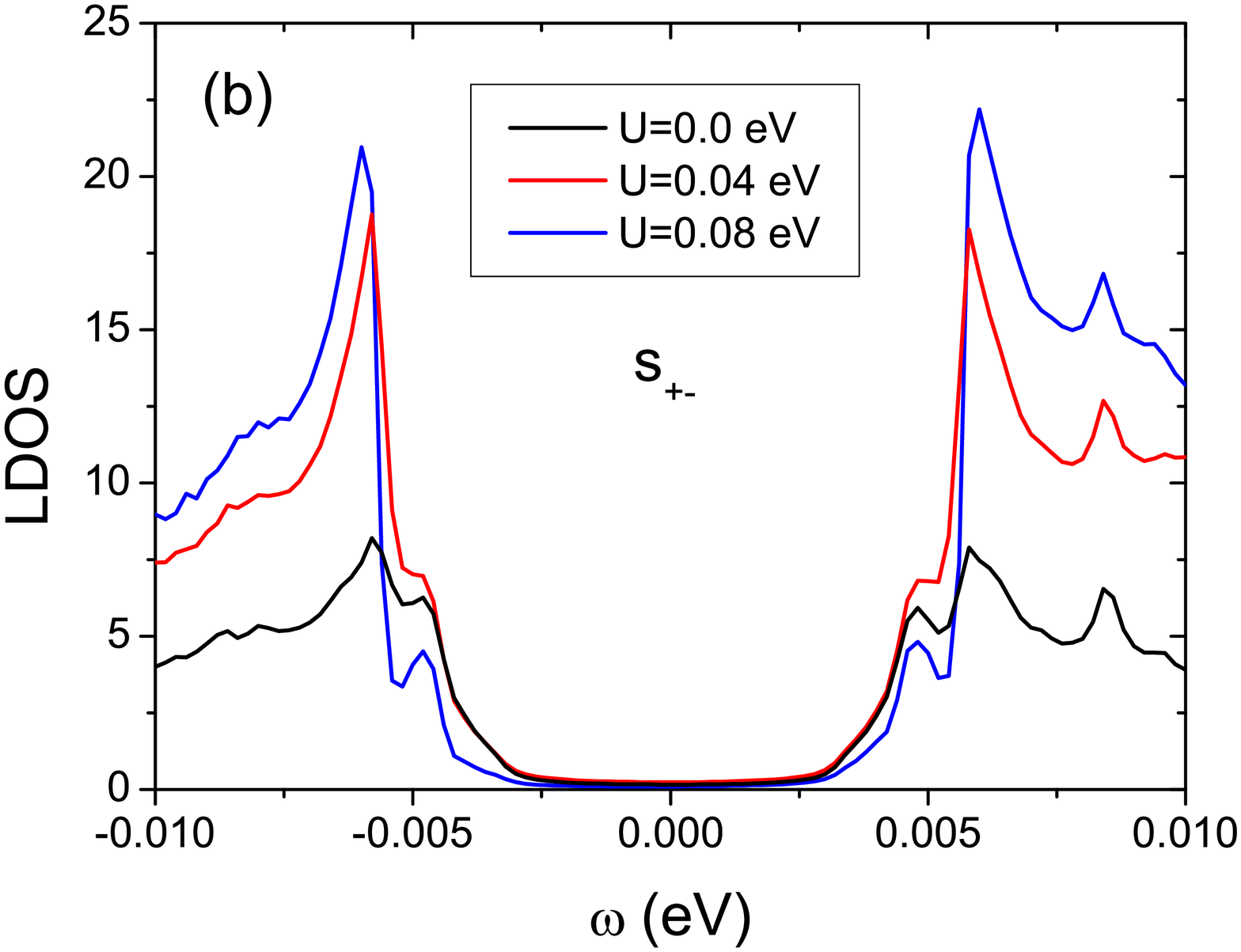}}
\rotatebox[origin=c]{0}{\includegraphics[angle=0,
           height=1.25in]{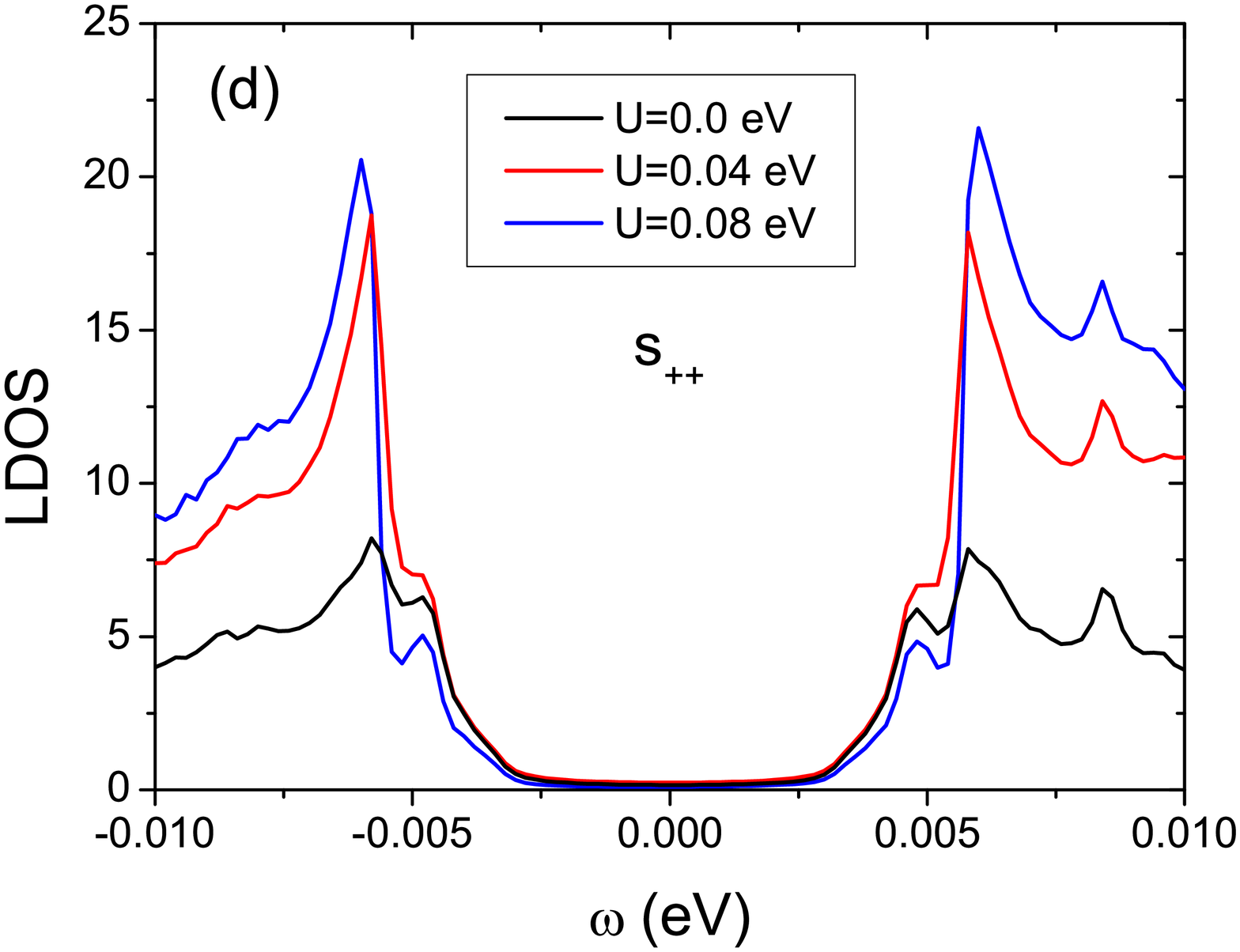}}
\caption {(Color online) The LDOS on the  nearest neighboring
Fe sites of the As (Te, Se) vacancy as a
function of the bias voltage $\omega$ under different $U$
at optimal electron doping ($15\%$)
for $s_{+-}$ pairing symmetry $
\Delta_{uv{\rm \bf k}}=\frac{1}{2}\Delta_0(\cos k_x+\cos k_y)$
in (a) and (b)
and $s_{++}$ pairing symmetry $
\Delta_{uv{\rm \bf k}}=\frac{1}{2}\Delta_0|\cos k_x+\cos k_y|$
in (c) and (d), respectively.
Here, $\Delta_0=5.8$ meV is the superconducting energy gap measured by
STM experiments.}
\end{figure}

The Hamiltonian describing a single As (Te, Se) vacancy located at point
$(0,\frac{1}{2})$ in Fe sublattice B or $(-\frac{1}{2},0)$ in Fe sublattice A
( see Fig. 1) can be written as $H=H_0+H_{\rm BCS}+H_V$,
where $H_0$ is the two-orbit four-band tight binding model proposed in
Ref. [10], $H_{\rm BCS}$ is the mean field BCS pairing Hamiltonian
in the Fe-Fe plane, $H_V= U\sum_{\alpha,\sigma}[c^+_{A,\alpha,(0,0),\sigma}
c_{A,\alpha,(-1,0),\sigma}+c^+_{B,\alpha,(0,1),\sigma}
c_{B,\alpha,(0,0),\sigma}+{\rm h.c.}]
+W\sum_{\alpha,\sigma}[c^+_{A,\alpha,(0,0),\sigma}c_{A,1-\alpha,(-1,0),\sigma}
+c^+_{B,\alpha,(0,1),\sigma}
\times c_{B,1-\alpha,(0,0),\sigma}+{\rm h.c.}] $,
$\alpha=0$ and $1$  represent  the degenerate orbitals $d_{xz}$ and $d_{yz}$,
respectively, $c^+_{A(B),\alpha,(i,j),\sigma }$
($c_{A(B),\alpha,(i,j),\sigma}$) creates (destroys) an $\alpha$
electron with spin $\sigma$ (=$\uparrow$ or $\downarrow$) in the unit cell
$(i, j)$ of the Fe sublattice A (B), and
$U$ ($W$) is the loca hopping correction between the same (different) orbitals
due to the As (Te, Se) vacancy. Because a vacancy cannot mix
$d_{xz}$ orbital and $d_{yz}$ orbital, we always have $W=-t_4$.

\begin{figure}
\rotatebox[origin=c]{0}{\includegraphics[angle=0,
           height=1.55in]{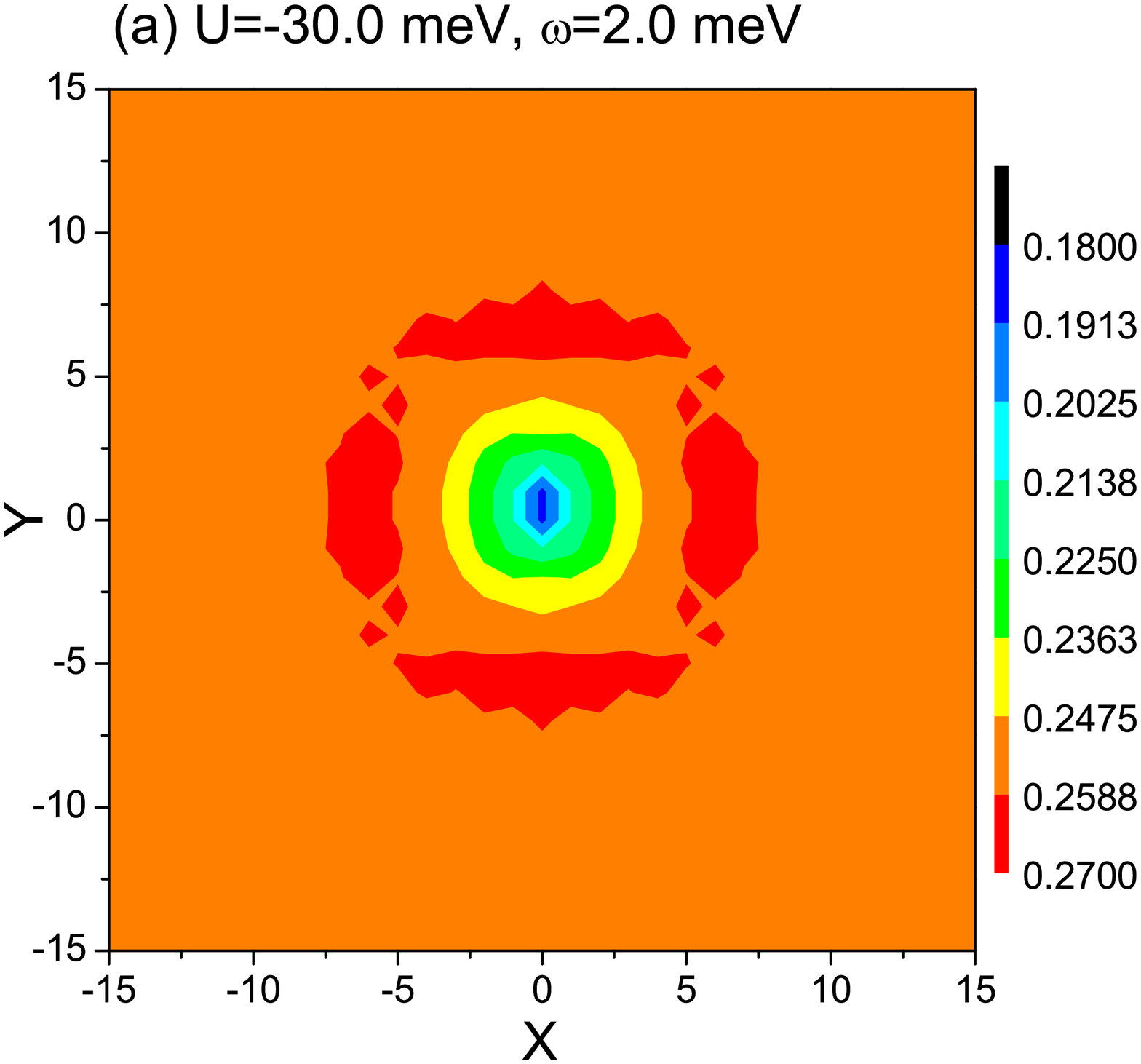}}
\rotatebox[origin=c]{0}{\includegraphics[angle=0,
           height=1.55in]{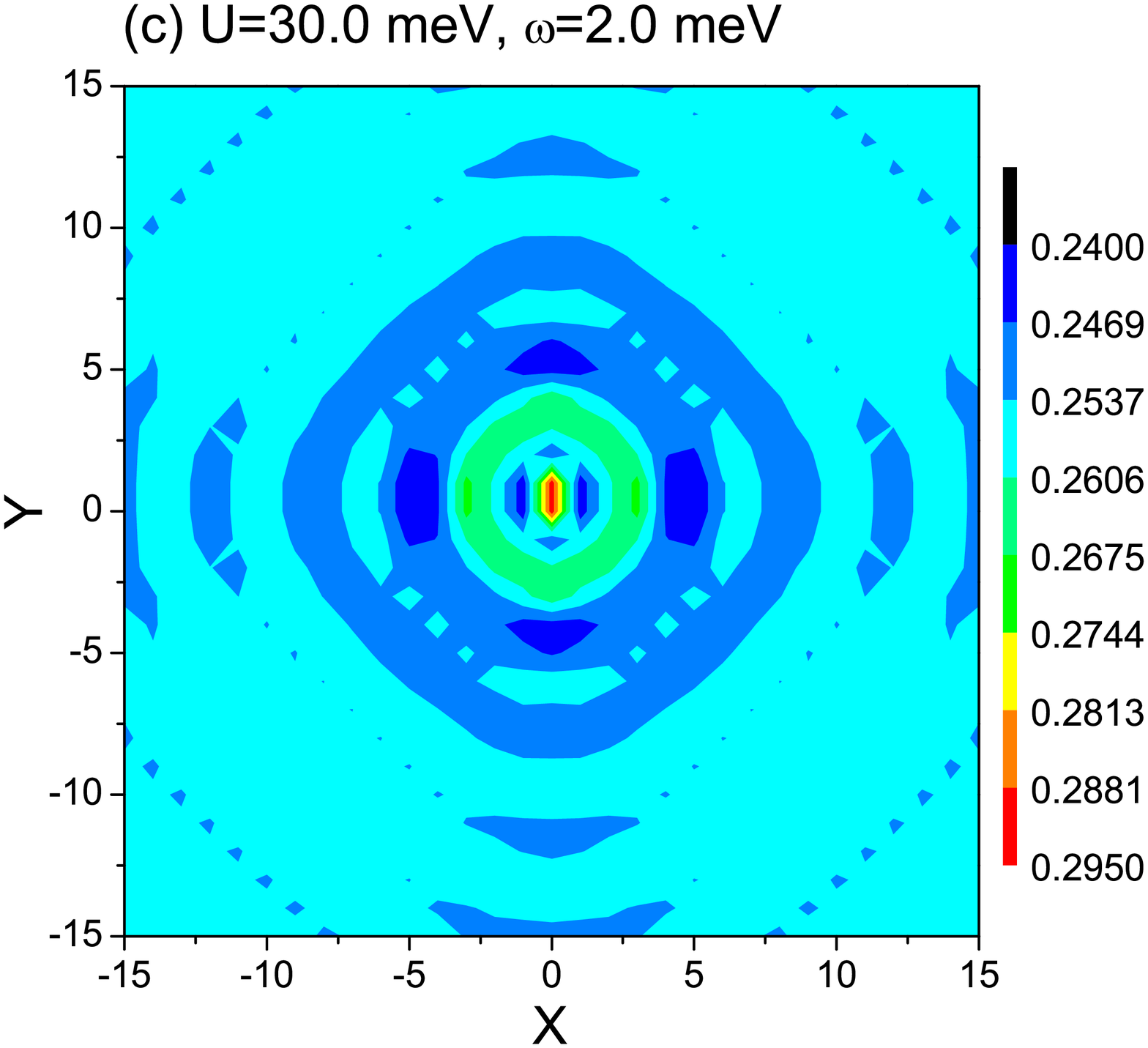}}
\rotatebox[origin=c]{0}{\includegraphics[angle=0,
           height=1.55in]{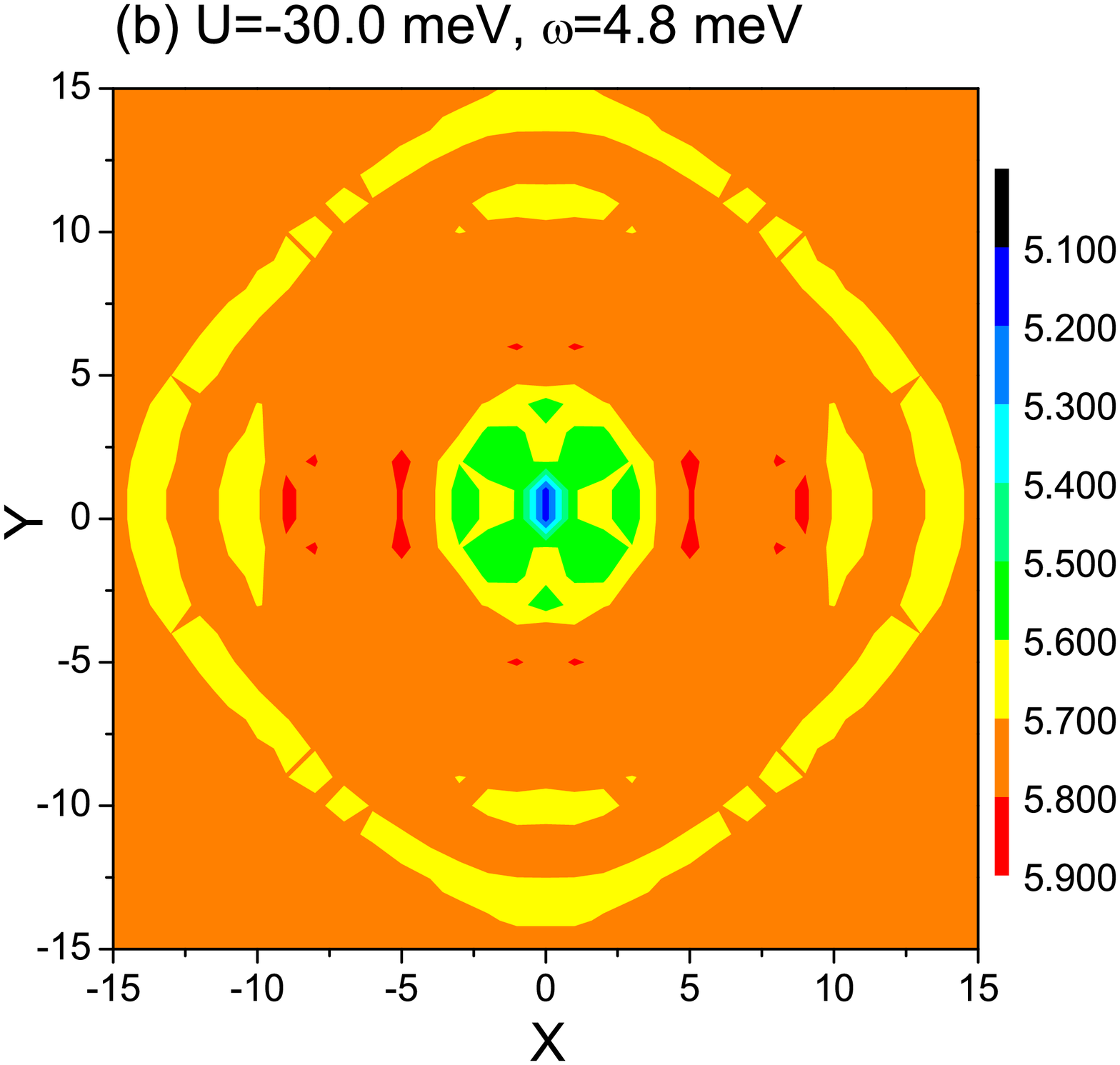}}
\rotatebox[origin=c]{0}{\includegraphics[angle=0,
           height=1.55in]{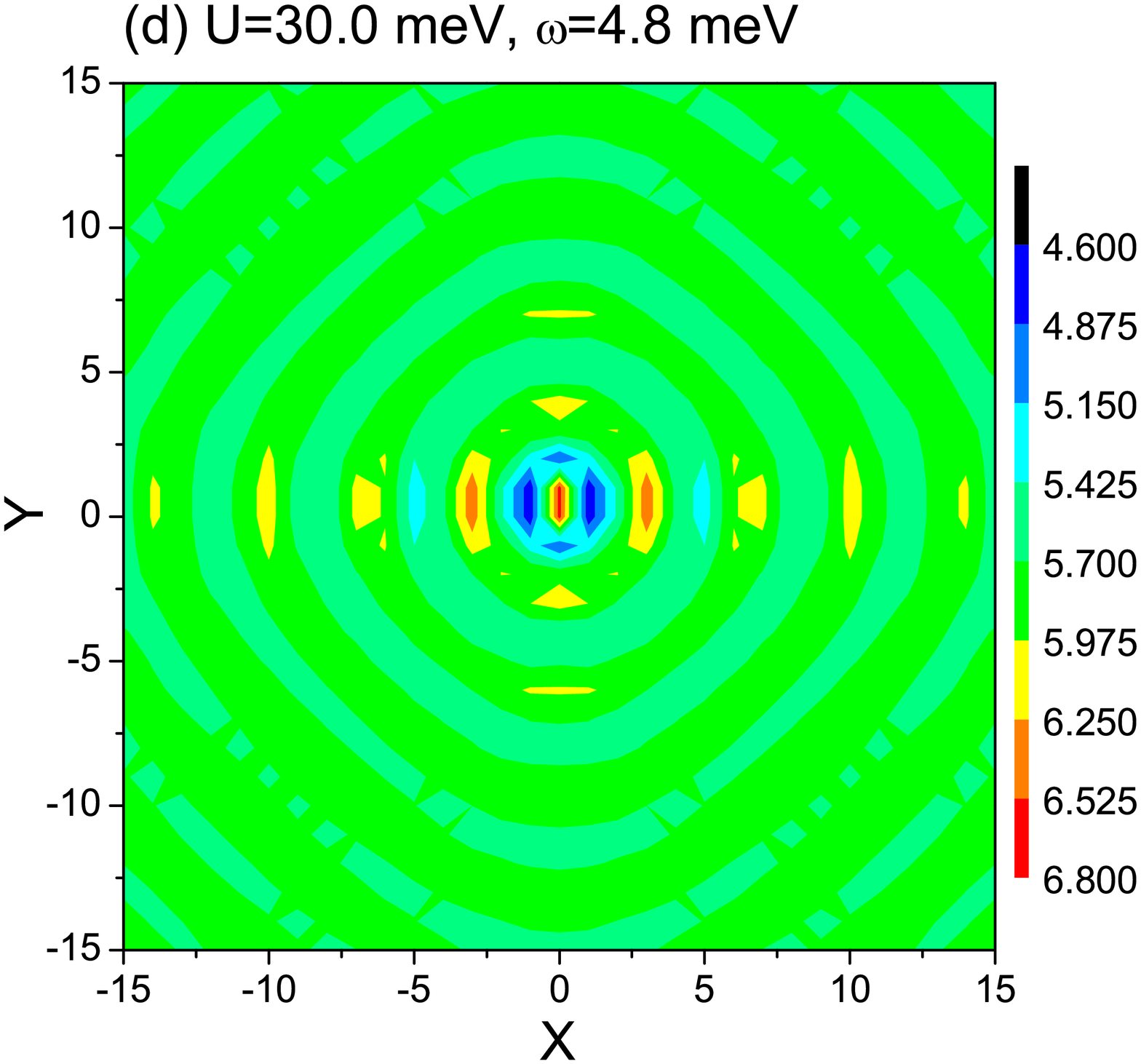}}
\caption {(Color online) The LDOS images near the As (Te, Se)
vacancy in the Fe sublattice $B$ under different $U$ and $\omega$
at optimal electron doping ($15\%$) for $s_{+-}$ pairing symmetry
$\Delta_{uv{\rm \bf k}}=\frac{1}{2}\Delta_0(\cos k_x+\cos k_y)$ with
$\Delta_0=5.8$ meV.}
\end{figure}

After introducing first the Fourier transformations
${c}_{A(B),\alpha,(i,j),\sigma}=\frac{1}{\sqrt{N}}\sum_{\bf
k}{c}_{A(B),\alpha,{\bf k},\sigma}e^{i(k_x x_i+k_y y_j)}$ with $N$ the
number of unit cells and the canonical transformations
for ${c}_{A,\alpha,{\bf k},\sigma}$ and ${c}_{B,\alpha,{\bf k},\sigma}$,
and then taking the Bogoliubov transformations for new
fermion operators, we can solve exactly the Hamiltonian $H$ for
a single ligand vacancy in iron-based superconductors
by using the T-matrix approach [10,29,34].
The analytic formulas for the Green's functions in momentum space
have been derived. The LDOS on the Fe sites at different bias voltages
and the Fourier component of LDOS (FCLDOS) can be obtained
through the Green's functions. Here we have calculated a square Fe lattice
with $N=500\times 500$ unit cells, which is enough to ensure the
accuracy of theoretical results.
We have also employed the energy band parameters:
$t_1=-0.5$ eV, $t_2=-0.2$ eV, $t_3=1.0$ eV,
and $t_4=-0.02$ eV, which are same with the previous works
[10,23,26,27,29,30,33]. We note that the electron-doped
BaFe$_{2-x}$Co$_x$As$_2$ has a superconducting energy gap
$\Delta_0=5.8$ meV observed by the STM experiments [8,9,35].
After a lots of numerical calculations, we found that
when $U>0.12$ eV or $U<-0.22$ eV, the LDOS is negative at
some bias voltages, which is unphysical. In other words,
An As (Te, Se) vacancy can produce a local hopping modification
in the interval [-0.22, 0.12] eV. This manifests that
the ligand ions play an important role in forming
high temperature superconductivity in iron-based superconductors.

We plot the LDOS on the point (0,0) or (0,1) in the Fe sublattice B
as a function of the bias voltage
$\omega$ under different $U$ and the optimal electron doping
(15$\%$) for the $s_{+-}$ pairing symmetry
$\Delta_{uv{\rm \bf k}}=\frac{1}{2}\Delta_0(\cos k_x+\cos k_y)$
in Figs. 2(a) and 2(b) and the $s_{++}$ pairing symmetry
$\Delta_{uv{\rm \bf k}}=\frac{1}{2}\Delta_0|\cos k_x+\cos k_y|$
in Fig. 2(c) and 2(d), respectively. Here, $u=0 (1)$ represents
the Fermi surfaces around M ($\Gamma$) point while $v=0 (1)$
denotes the outer (inner) Fermi surfaces around M or ($\Gamma$)
point [10].
Obviously, the curves of the LDOS for the $s_{+-}$ pairing symmetry
coincide completely with those for  the $s_{++}$ pairing symmetry
when all the other parameters are fixed.
Two resonance peaks exhibit in the LDOS at
$\omega=\pm 4.8$ meV and their locations don't move with increasing
$U$. These results agree well the recent STM observations [8,9].
However, even if $U=0$, two resonance peaks still stand
there. This means that the hybridization among Fe $d$ and
As (Te, Se) $p$ orbitals cannot be neglected.
When $U<0$, with decreasing $U$, the superconducting coherence peaks
and the resonance peaks are suppressed, and their heights at positive
bias voltages become lower than those at negative bias voltages.
It is quite obvious that the resonance peak is higher than the coherence
peak at the positive energy side for enough small $U$ [see Fig. 2(a)]. 
If $U>0$, with increasing $U$, the superconducting coherence peaks
grow up, but the resonance peaks first become higher, then become
lower. However, the coherence peaks and the resonance peaks are symmetric
with respect to the bias voltage in Fig. 2(b).
We note that the in-gap resonance peaks is irrelative to the phase of
the superconducting order parameter, similar to those induced by interstitial
excess Fe impurities in iron-based superconductor Fe(Te, Se) [28-30].
Therefore, very different from nonmagnetic impurities on the Fe sites [10],
such a ligand vacancy cannot be used to distinguish the 
$s_{+-}$ and $s_{++}$ pairing symmetries.

\begin{figure}
\rotatebox[origin=c]{0}{\includegraphics[angle=0,
           height=1.2in]{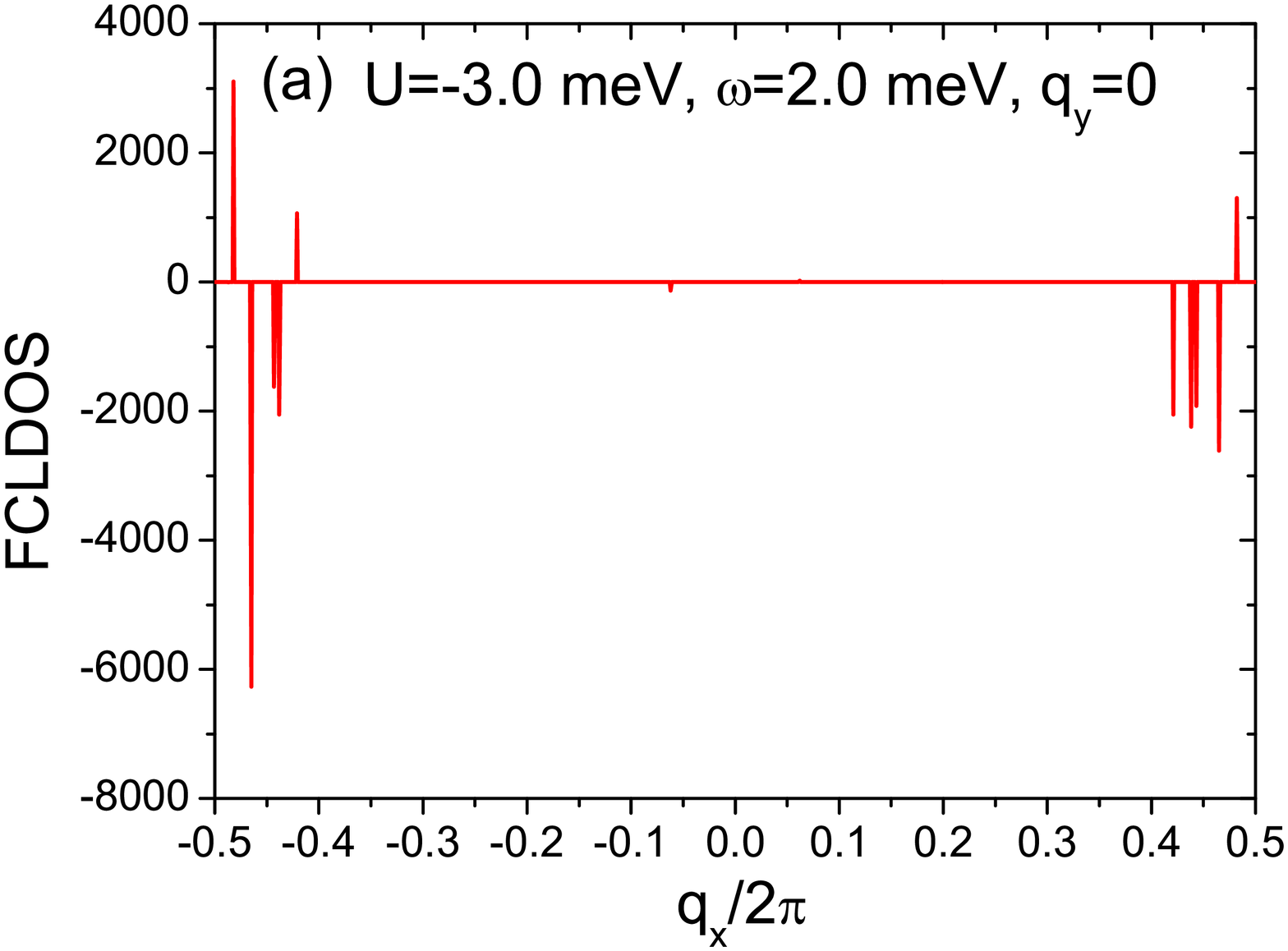}}
\rotatebox[origin=c]{0}{\includegraphics[angle=0,
           height=1.2in]{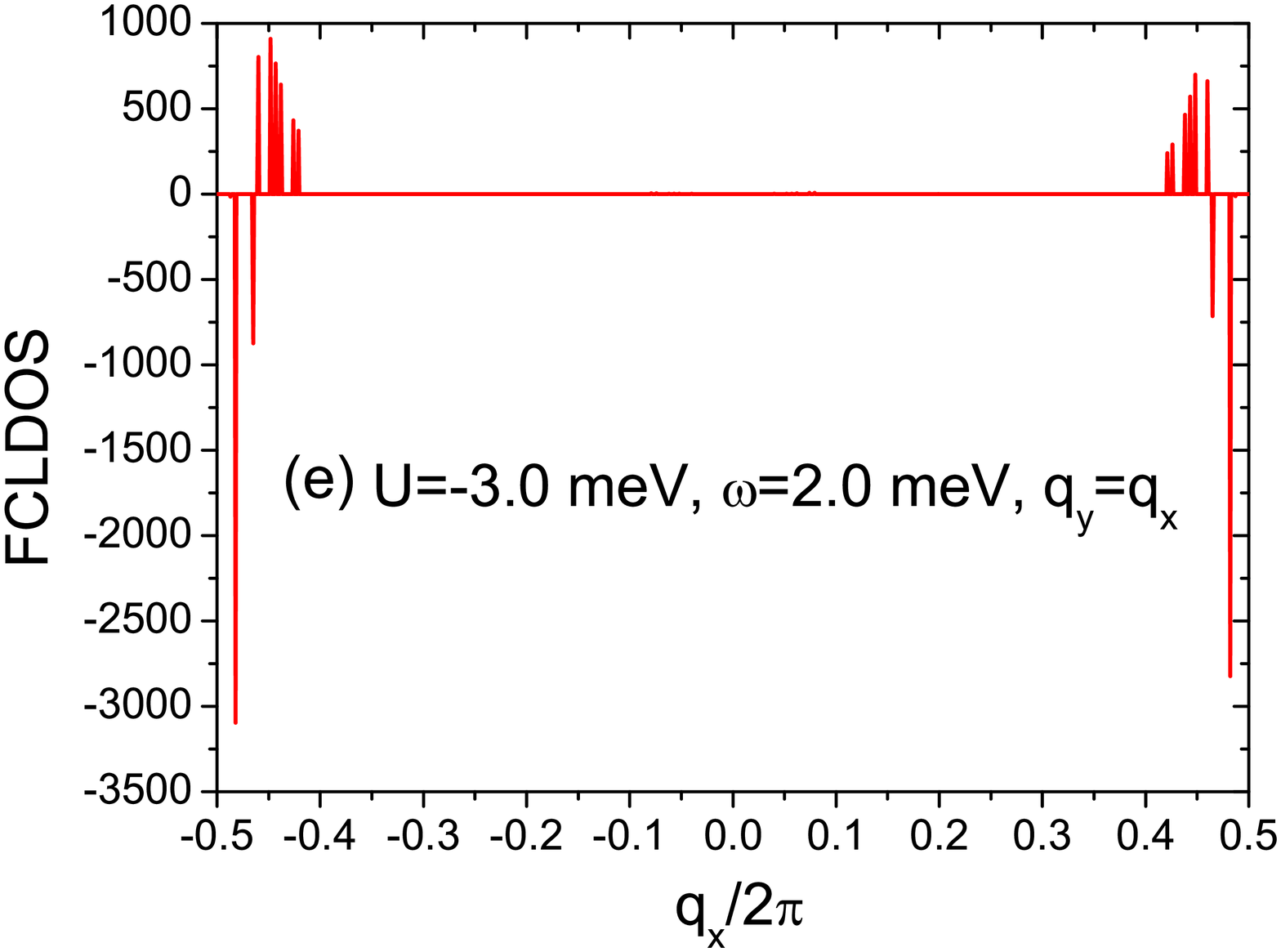}}
\rotatebox[origin=c]{0}{\includegraphics[angle=0,
           height=1.2in]{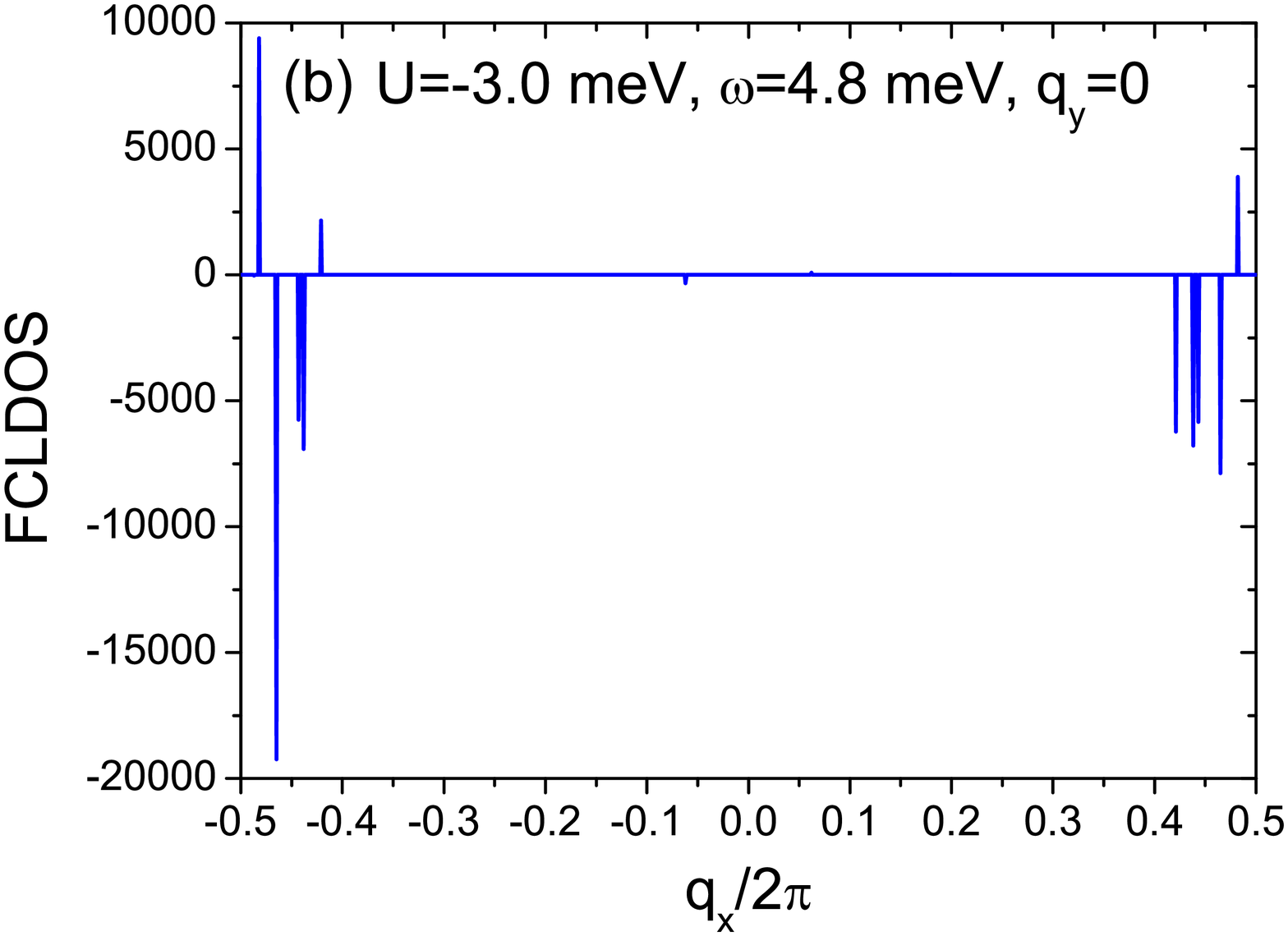}}
\rotatebox[origin=c]{0}{\includegraphics[angle=0,
           height=1.2in]{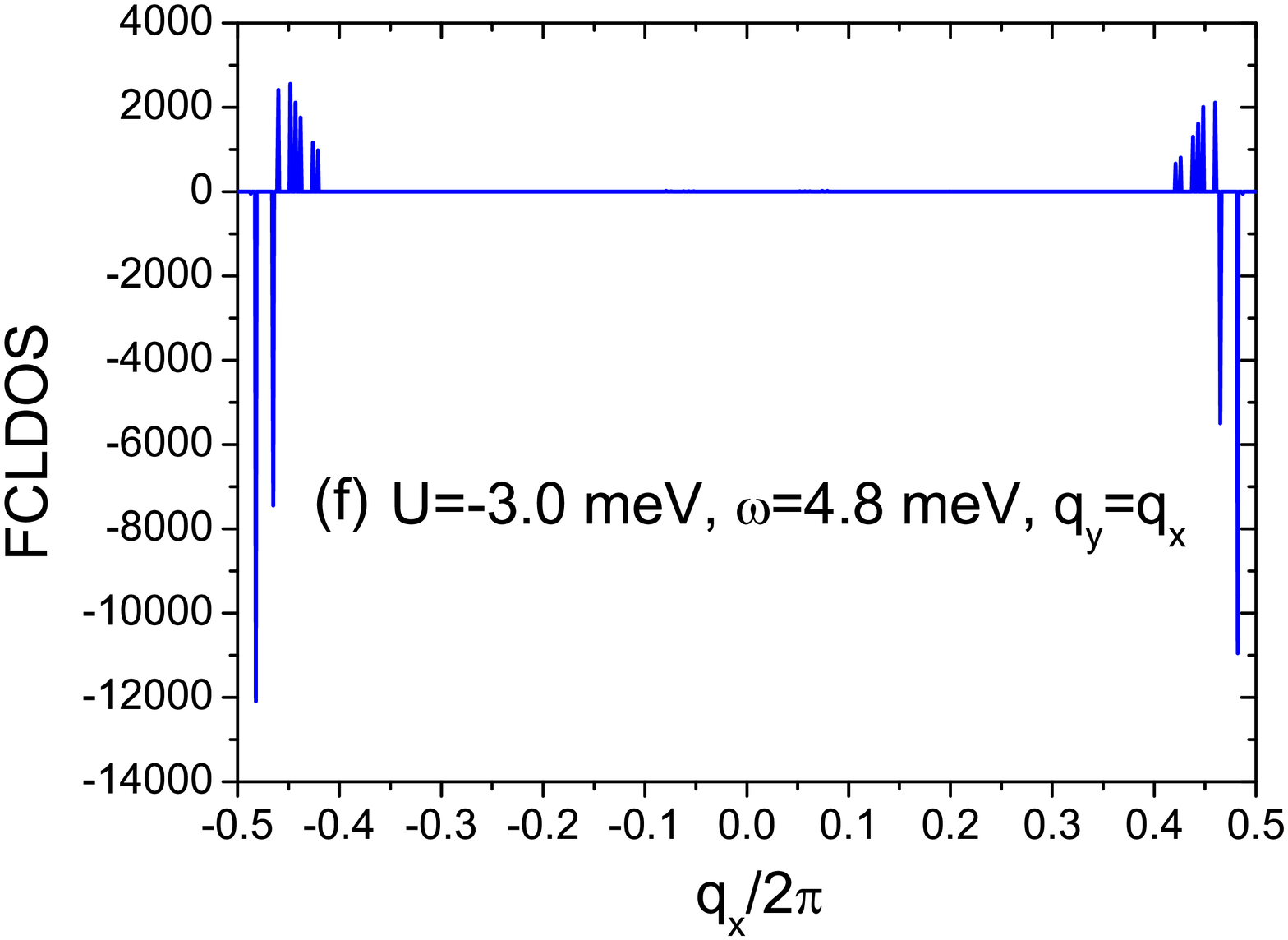}}
\rotatebox[origin=c]{0}{\includegraphics[angle=0,
           height=1.2in]{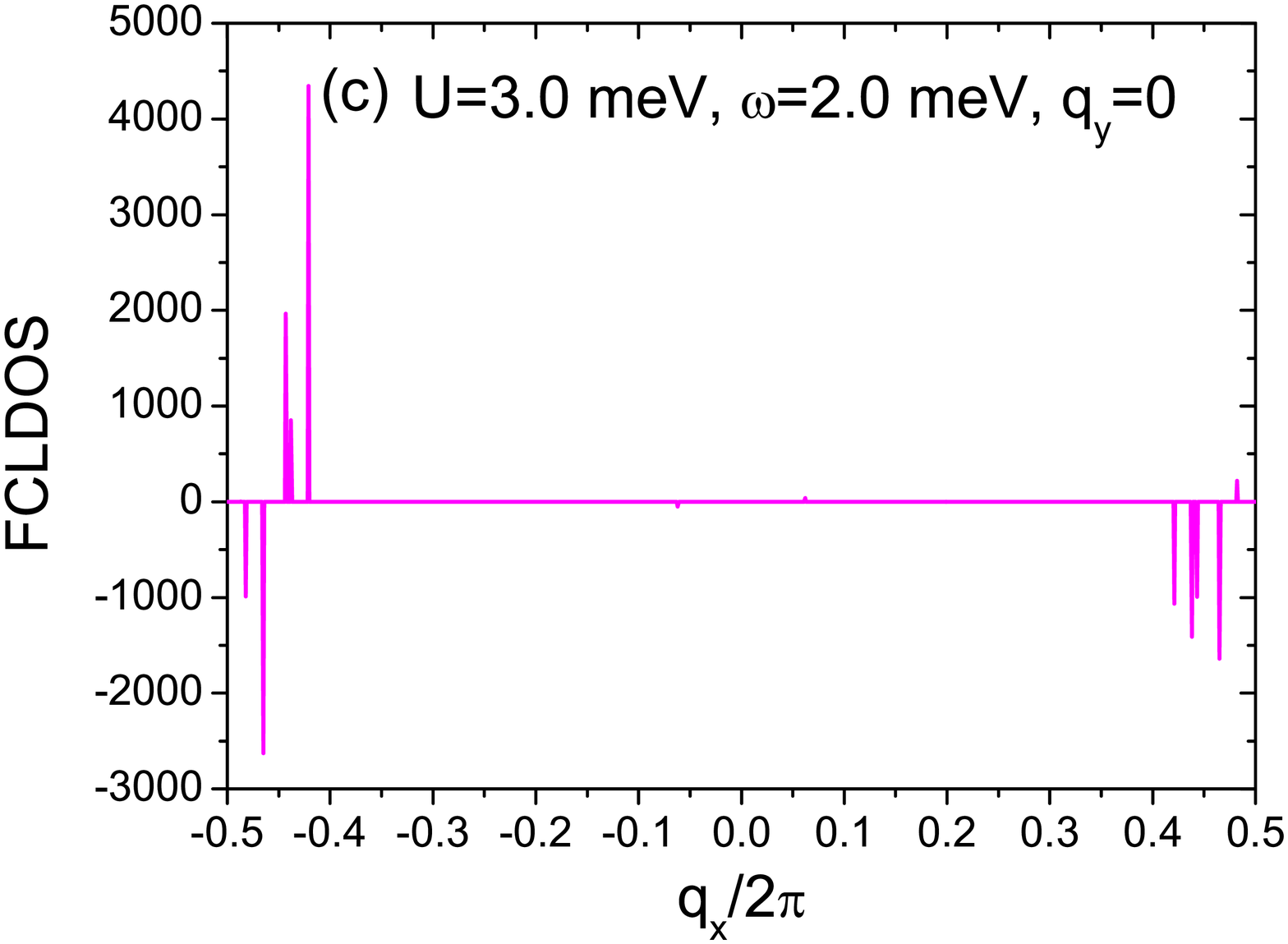}}
\rotatebox[origin=c]{0}{\includegraphics[angle=0,
           height=1.2in]{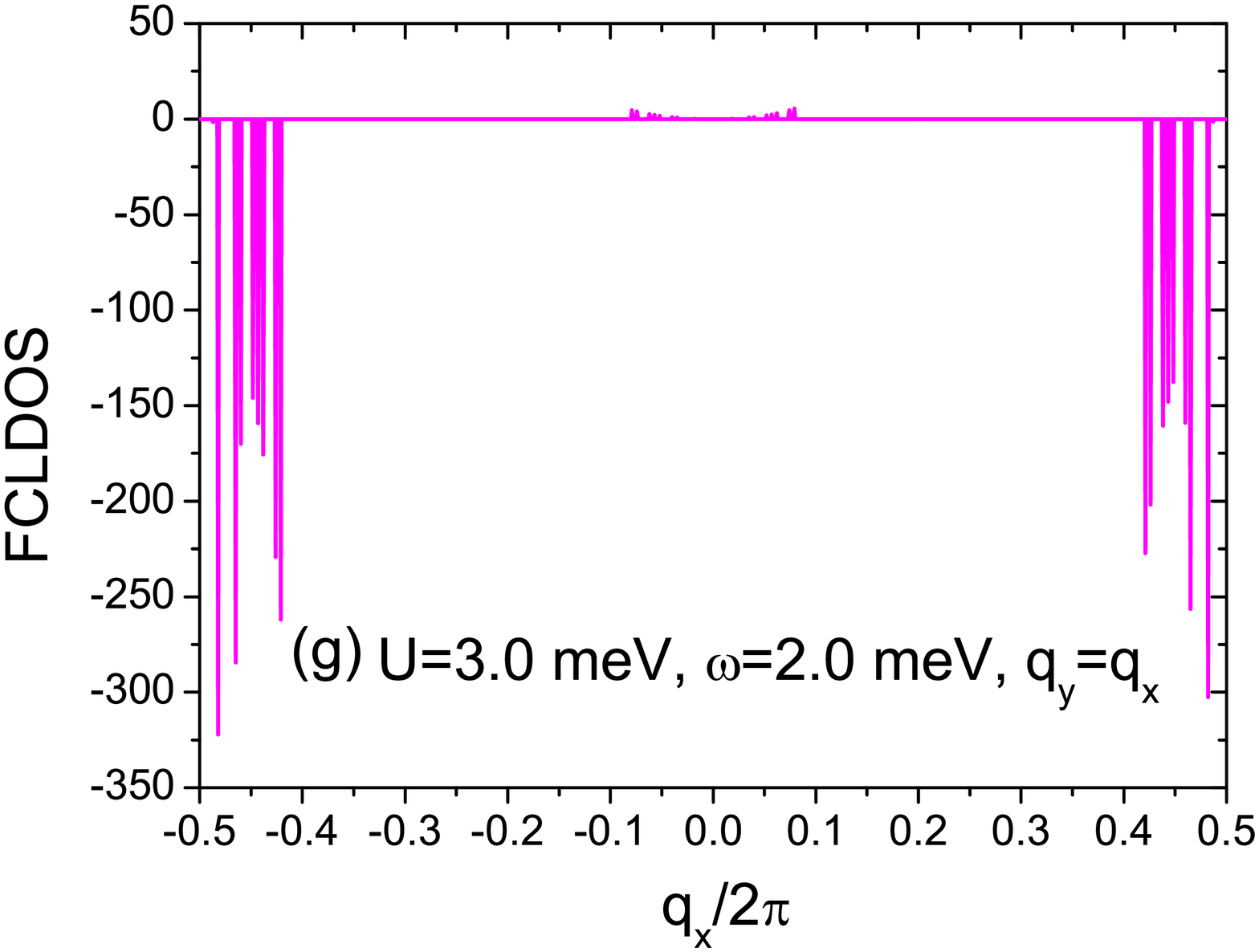}}
\rotatebox[origin=c]{0}{\includegraphics[angle=0,
           height=1.2in]{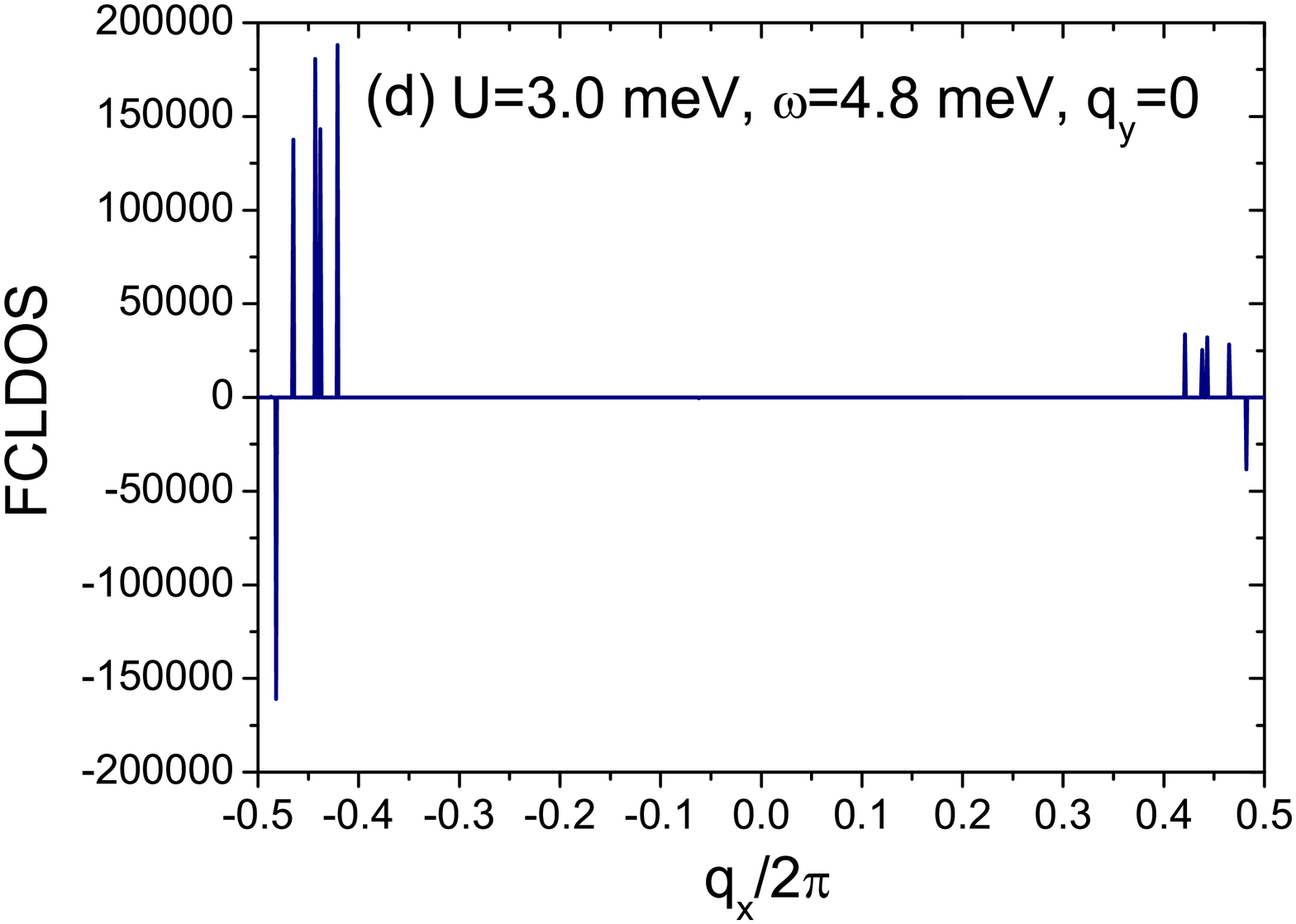}}
\rotatebox[origin=c]{0}{\includegraphics[angle=0,
           height=1.2in]{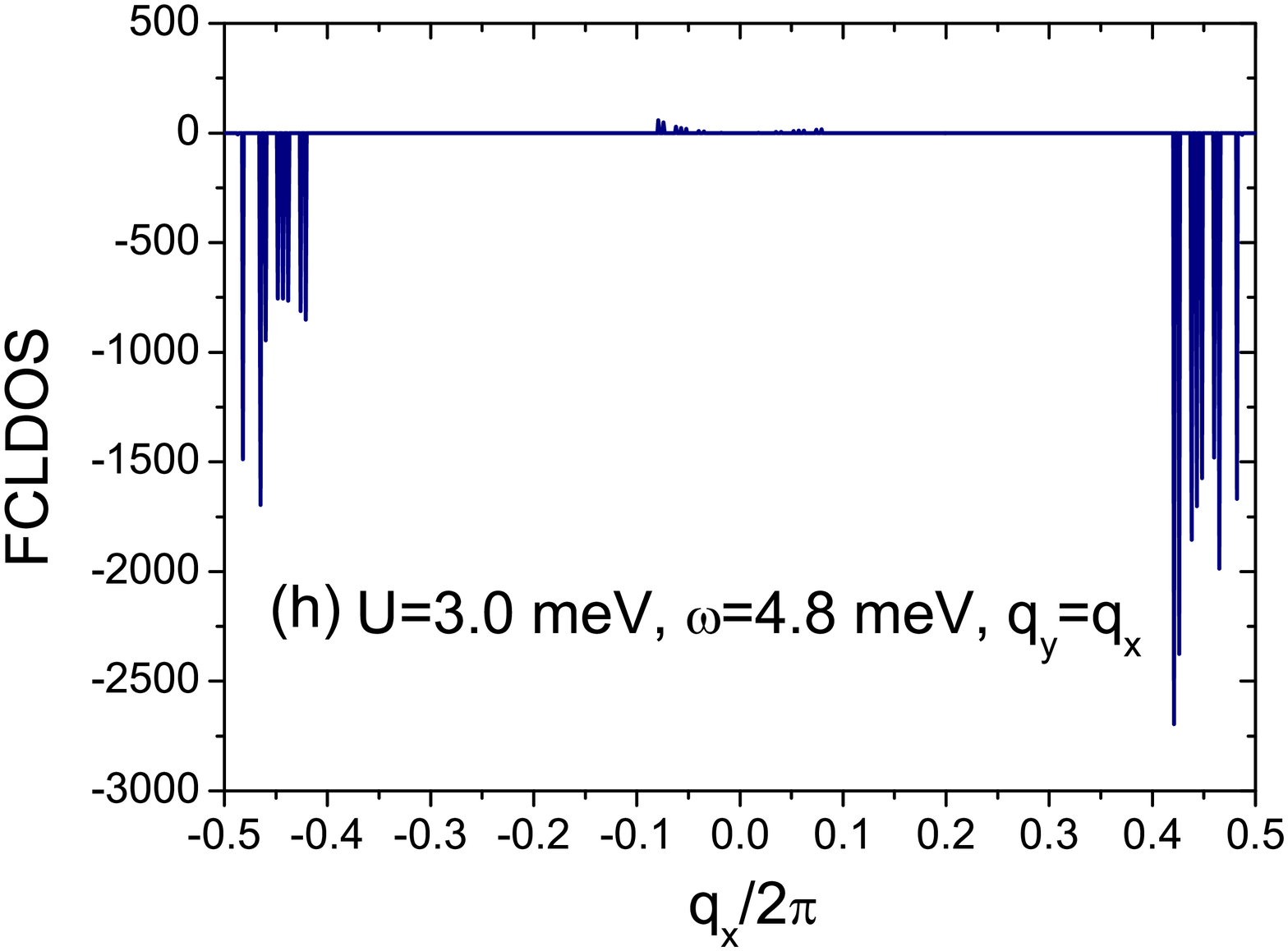}}
\caption {(Color online) The FCLDOS along $(\pi, 0)$ direction
in (a)-(d) and $(\pi, \pi)$ direction in (e)-(h)
under different $U$ and $\omega$
at optimal electron doping ($15\%)$
for the $s_{+-}$ pairing symmetry $
\Delta_{uv{\rm \bf k}}=\frac{1}{2}\Delta_0(\cos k_x+\cos k_y)$ with
$\Delta_0=5.8$ meV.}
\end{figure}

Fig. 3 shows the LDOS images at different $U$ and $\omega$ under optimal
electron doping for the $s_{+-}$ pairing symmetry
in the Fe sublattice B with $31\times 31$ sites due to quasiparticle interference.
The As (Te, Se) vacancy is located at the point $(0, \frac{1}{2})$.
Because up As (Te, Se) and down As (Te, Se) are inequivalent
in the surface layer, all the LDOS images have a $C_2$ symmetry.
Obviously, the LDOS on the points $(0,0)$ and $(0,1)$ has a maximum
(minimum) value for $U>0 (<0)$. Hence we can judge that the As (Te, Se)
vacancy is attractive or repulsive according to the extreme value of the LDOS
on the nearest neighboring Fe sites around it.
When $U=-30.0$ meV, the LDOS has $0^o$ modulation at $\omega=2.0$ meV
in Fig. 3(a). With increasing $\omega$, both $0^o$ and $45^o$ stripes show
up in Fig. 3(b). If $U=30.0$ meV, the real space LDOS also possesses
energy-dependent modulations along both $0^o$ and $45^o$ directions
at $\omega=2.0$ meV and 4.8 meV [see Figs. 3(c) and 3(d)].

To understand the origin of the LDOS modulations
produced by quasiparticle interference and the modulation periods,
we have obtained the FCLDOS at different $U$ and $\omega$.
Figs. 4(a)-4(d) and Figs. 4(e)-4(h) exhibit the modulation wave
vectors and their intensities along $0^o$ and $45^o$ directions,
respectively, corresponding to the real space LDOS
images in Fig. 3. It is very interesting that all the modulation
wave vectors are independent of $U$ and $\omega$. Therefore,
the energy-dependent charge modulations are due to the
variations of the FCLDOS intensities at the modulation
wave vectors with energy. Comparing carefully Figs. 4(e)-4(h) with
Figs. 4(a)-4(d), we can see clearly that the modulation
wave vectors along $0^o$ direction are nothing but the
$x$ components of those along $45^o$ direction.
The magnitudes of the modulation wave vectors along $45^o$ direction
mainly distribute in the range of $0.84\sqrt{2}\pi\sim
0.97\sqrt{2}\pi$. Because the single vacancy is not located 
at the center of the Fe sublattice B, the
modulation wave vectors have fine structures with double
lines, which are never reported.
Obviously, this quasiparticle interference phenomenon is different
from that in the cuprate superconductors, where
the charge modulation wave vectors shift with the bias voltage,
and their change trends along $0^o$ (antinodal) and $45^o$ (nodal)
directions are just opposite with increasing energy [36-38].

\begin{figure}
\rotatebox[origin=c]{0}{\includegraphics[angle=0,
           height=2.6in]{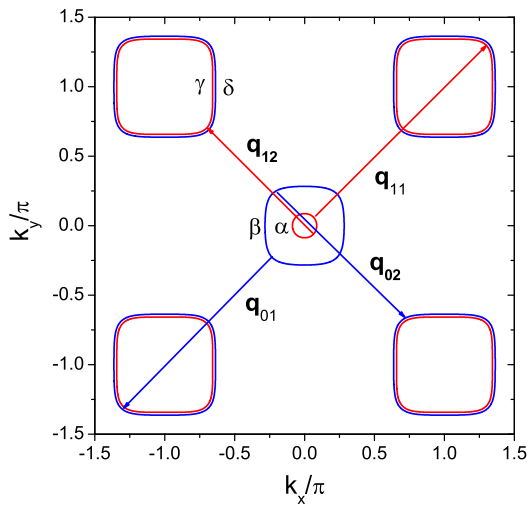}}
\caption {(Color online) The Fermi surfaces of iron-based superconductors
at optimal electron doping ($15\%)$ and the allowed nesting vectors.}
\end{figure}

Now we determine the modulation wave vectors in the LDOS images.
Fig. 5 depicts two hole Fermi surfaces (i.e. $\alpha$ band with
$u=v=1$ and $\beta$ band with $u=1$ and $v=0$) around $\Gamma$ point
and two electron Fermi surfaces (i.e. $\gamma$ band with
$u=0$ and $v=1$ and $\delta$ band with $u=v=0$) around M point
of iron-based superconductors at optimal electron doping [10].
According to the analytic expression of the Green's function
or the LDOS, we found that the interband transition is only
allowed for those bands with the same index $v$.
In Fig. 5, ${\bf q}_{0\tau}$ (${\bf q}_{1\tau}$) ($\tau=1$ and 2)
represent the allowed
nesting vectors connecting to two outer (inner) Fermi surfaces
around $\Gamma$ and M points, and
$|{\bf q}_{11}|> |{\bf q}_{01}|>|{\bf q}_{02}|> |{\bf q}_{12}|$.
We analyze in detail the numerical values of the modulation wave vectors
in Fig. 4 and the nesting vectors in Fig. 5, and conclude firmly
that the nesting vectors ${\bf q}_{v\tau}$ and their
vector differences ${\bf q}_{v1}-{\bf q}_{v2}$
are nothing but the the modulation wave vectors
in the LDOS images.

In summary, we have explored the impact of a single
As (Te, Se) vacancy on the electronic states in iron-based superconductors.
The ligand vacancy can induce two
robust resonance peaks in the superconducting energy gap
at the fixed symmetric positions about zero energy,
which are consistent with the STM experiments.
The resonance peaks are independent of the phase of the superconducting 
order parameter in the bulk, similar to zero energy
bound state produced by the interstitial Fe ions.
Because a magnetic field that is not too strong changes predominately
the phase of the superconducting order parameter,
we predict that these two bound states keep unchanged with
increasing the magnetic field strength, which could be detected
by STM experiments. The energy-dependent LDOS images possess 
$0^o$ and $45^o$ stripes with multiple periods.
The modulation wave vectors come from the nesting vectors of the Fermi
surfaces, which are independent of the local hoping correction 
and the bias voltage. The quasiparticle interference patterns 
induced by As (Te, Se) vacancies are undoubtedly originated 
in the nesting effect in iron-based superconductors.

The authors would like to thank Jiaxin Yin and Ang Li for useful discussions.
This work was supported by the Sichuan Normal University, the "Thousand
Talents Program" of Sichuan Province, China, the Texas Center
for Superconductivity at the University of Houston,
and by the Robert A. Welch Foundation under grant No. E-1146.


\end{document}